\documentclass[11pt,twoside,A4]{article} 
\usepackage{times,fancyhdr}
\usepackage[dvips]{graphicx}
\usepackage{latexsym} 
\usepackage[affil-it]{authblk}
\usepackage{amsmath}
\usepackage{amssymb}
\usepackage{setspace}

\usepackage[top=4cm, bottom=4cm, left=4cm, right=4cm]{geometry}

\def\beq{\begin{equation}}
\def\eeq{\end{equation}}
\def\beqn{\begin{eqnarray}}
\def\eeqn{\end{eqnarray}}

\begin{document}
 
\title{A relativistic acoustic metric for planar black holes}
\author{Sabine Hossenfelder\thanks{hossi@nordita.org}} 
\affil{\small Nordita\\
KTH Royal Institute of Technology and Stockholm University\\
Roslagstullsbacken 23, SE-106 91 Stockholm, Sweden}
\date{}
\maketitle
\vspace*{-1cm}
\begin{abstract}
We demonstrate here that the metric of a planar black hole in asymptotic Anti-de Sitter 
space can, on a slice of dimension $3+1$, be reproduced as a relativistic acoustic metric. 
This completes an earlier calculation in which the non-relativistic limit was used, and also
serves to obtain a concrete form of the Lagrangian.
\end{abstract}

\section{Introduction}
 
The AdS/{\sc CFT} duality 
\cite{Maldacena:1997re,Witten:1998qj,Gubser:2002tv} identifies a gravitational
theory in asymptotic anti-de-Sitter (AdS) space with a strongly 
coupled conformal field theory ({\sc CFT}) on the boundary of the same space. This identification
can be used to translate a non-perturbative computation from a
strongly coupled condensed matter system to 
semi-classical gravity, and thereby make it conceptually easier to treat. This
approach has shown much promise to deal with systems such as the
quark gluon plasma and strange metals \cite{Hartnoll:2008kx,Hartnoll:2009sz,Horowitz:2010gk}, whose behavior is very
difficult to calculate by other methods.

In a previous paper \cite{Hossenfelder:2014gwa}, we pointed out the
possibility to use the AdS/{\sc CFT} duality to arrive at a new type of duality
which connects a strongly with a weakly coupled condensed matter system.
This can be done by combining AdS/{\sc CFT} with analog gravity.
Analog
gravity \cite{Unruh:1980cg, Barcelo:2005fc} is a way to assign an effective metric to certain types of weakly
coupled condensed matter systems. It can be shown that in these systems perturbations
propagate in the matter background according to an equation of motion 
identical to that of particles traveling in curved space. Gravitational analogs are known to exist for the Schwarzschild black hole \cite{Visser:1997ux,Garay:1999sk,Barcelo:2000tg} and expanding Friedmann-Robertson-Walker space-times that mimic
inflation in the early universe \cite{Volovik:2000ua,Weinfurtner:2004mu,Jain:2007gg,Lin:2012ft,Bilic:2013qpa}. 
Research in this area is presently very active, and theory and experiments both are being
rapidly developed \cite{Weinfurtner:2010nu,ex1}.

The effective metric 
is obtained from the behavior of the
background field, and completely specified by the degrees of freedom of the
condensed matter system. If this effective metric is also a gravitational dual of
a strongly coupled system, then this leads to a relation between the two
condensed matter systems that give rise to the same metric. The connection
between analog gravity and the AdS/{\sc CFT} duality was also explored in 
\cite{Bilic:2014dda,Das:2010mk,holo,holo2,Chen:2012uc}.

Not all metrics can be obtained as effective analog metrics from condensed
matter systems. It must both be possible to bring the metric into a specific
form, which amounts to a certain gauge condition, and it must then be
demonstrated that the condensed matter system needed to obtain this
metric does fulfill the equations of motion. In the previous paper \cite{Hossenfelder:2014gwa}
it was demonstrated that the type of metrics used to model strongly coupled systems 
via the AdS/{\sc CFT} duality can indeed also be obtained
as effective metrics of a weakly coupled condensed matter system. 

While intriguing evidence, this finding by itself does not suffice to show that there is a new duality between
weakly and strongly coupled condensed matter systems, because the
identification of the metric used in the AdS/{\sc CFT} correspondence as an
effective metric does only demonstrate that the semi-classical limit is
identical. It does however show that this necessary condition 
is fulfilled and thus represents a first step on the way to a more general
proof.

In the present paper, we will look at the next step, which is to demonstrate that there exists a relativistic
completion of the system used in  \cite{Hossenfelder:2014gwa}. In \cite{Hossenfelder:2014gwa} it was found that the
non-relativistic limit is good towards the boundary of AdS-space, but not close by the horizon. This 
is unfortunate because the near-horizon region is the most interesting part, so we will here derive a 
relativistic acoustic metric and recover the non-relativistic limit. We will find
that this also tells us more about the form of the Lagrangian of the analog
gravity system than could be extracted from the non-relativistic limit.

Throughout this paper we use units in which the speed of light and $\hbar=1$. 
The constant $c$ denotes the speed of sound and {\sl not} the speed of light.
The metric signature is $(-1,1,1,1)$. Small Greek indices run from zero to three. With non-relativistic we refer to the limit
in which the four-velocity is $\ll c$.

\section{Framing the question}

The most instructive way to obtain the effective analog metric of a fluid is to
use the Lagrangian approach in a mean-field approximation and then
derive the equations of motion for perturbations around the mean \cite{Barcelo:2005fc,Bilic:2013qpa,Hossenfelder:2014gwa}. It can
be shown then that the perturbations obey a wave-equation that is identical
to the wave-equation in a curved background whose metric is the effective
analog metric.  The form of the metric one obtains in this way depends on the 
Lagrangian, and we will be dealing here specifically with an 
effective metric known as the `acoustic metric' because it determines the
propagation of sound waves. There are other types of analog metrics
for different systems, for example the optic metric \cite{Barcelo:2005fc,Gordon,optic1,optic2}, but these will not be discussed here.

It must be emphasized that the key finding of analog gravity is not trivial. While the assumption
of Lorentz-invariance of the original Lagrangian 
already tells us that the resulting equation of motion for the perturbation must
respect this symmetry too, this alone does not single out a wave-equation in
curved space.
The important property of the resulting equation is that it separates the 
background from the perturbations in just the right way so that the degrees
of freedom of the background can be collected in something that takes the
form of a metric tensor, or that the derivatives can be reformulated as
covariant derivatives in curved space respectively. Just by looking at all the terms that are possible 
when one
requires that indices are contracted suitably, there could
be combinations between the background field and the perturbations that do
not lend themselves to the description in terms of an effective metric.

Indeed, it is interesting to observe that Lorentz-invariance in the equations governing
the propagation of the exitations can emerge even if the equations of the background
themselves are in the non-relativistic limit \cite{Fagnocchi:2010sn}. The Lorentz-group that is relevant here
is that in which the limiting velocity is the speed of sound, and not the speed of
light. At high energies, this emergent Lorentz-invariance can be violated, a phenomenon
that has been used to study the robustness under UV-corrections of quantum
field theory in curved background \cite{Liberati:2005id}. 

Concretely, we take a Lagrangian for a real scalar field $\theta$ of the form
\beqn
{\cal L} = {\cal L}(\chi (\partial \theta) - V )~. \label{L}
\eeqn
which depends on a kinetic energy term
\beqn
\chi = \eta^{\mu\nu} \left( \partial_\nu \theta \right) \left( \partial_\mu \theta \right)
\eeqn
and some additional potential $V$ that may include a mass term (more about the potential later).  $\eta$ is the metric in the laboratory
that hosts the analog gravity system. We assume that this space-time in the laboratory is flat, ie that its curvature tensor vanishes. However,
since we are free to choose a coordinate system, the metric might not have the Minkowski-form.

We then
make a perturbation around a background field that is assumed to fulfill the equations of motion, $\theta= \theta_0 + \varepsilon \theta_1$. The inverse of the
analog metric in terms of derivatives with respect to the field then takes the rather simple form \cite{Barcelo:2005fc,Bilic:1999sq,Bilic:2013qpa,Hossenfelder:2014gwa}
\beqn
\sqrt{-g} g^{\mu \nu} = - \sqrt{-\eta} \frac{\partial^2 {\cal L}}{\partial (\partial_\nu \theta_0) \partial (\partial_\mu \theta_0)} ~. \label{geff}
\eeqn
Here and in the following, the lower index $0$ refers to quantities describing the background field (at zeroth order). 
This metric then has to be rewritten into the hydrodynamic variables of the background fluid, and be inverted.

One can identify the pressure $p_0$, the density $\rho_0$ and the fluid-velocity $u_\nu$ by comparing the stress-energy
derived from the Lagrangian (\ref{L}) to the familiar stress-energy tensor of a fluid, from which one finds 
\beqn
u_\nu = \frac{\partial_\nu \theta_0}{\sqrt{\chi}}~,~p_0 = {\cal L}~,~\rho_0 = 2 \chi \frac{\partial {\cal L}}{\partial \chi} - {\cal L}~. \label{ident}
\eeqn
To match this to the notation of \cite{Visser:2010xv}, it is $\chi = (\rho_0+p_0)^2/n_0^2$, where $n_0$ is the particle-density of the fluid
and $\chi$ is the specific enthalpy. The four-velocity is normalized to minus one
\beqn
\eta^{\mu\nu}u_\mu u_\nu = - 1~. \label{norm}
\eeqn
With these variables one then gets the acoustic metric
\beqn
g_{\mu \nu} =  \left(  \frac{\rho_0+p_0}{c \chi} \right)  \left( \eta_{\mu\nu} + \left(1- c^2 \right) u_\mu u_\nu \right) ~, \label{acmet}
\eeqn
where $c$ is the speed of sound and defined by
\beqn
\frac{1}{c^2} = \frac{\partial \rho_0}{\partial p_0}  = \frac{2 \chi \partial^2 {\cal L}/\partial \chi^2 + \partial {\cal L}/\partial \chi}{\partial{\cal L}/\partial \chi} ~. \label{sos}
\eeqn

What we aim to show here is that there exists a scalar field Lagrangian of the general form  (\ref{L}) that gives rise to an acoustic metric which describes the planar black hole 
in asymptotic $4+1$ dimensional AdS that reads
\beqn
{\rm d}s^2 = - \frac{L^2}{ z^2}  \gamma(z)  {\rm d} {\tilde t}^2 + 
\frac{L^2}{z^2} \gamma( z)^{-1} {\rm d} z^2 + \frac{L^2}{z^2}  \sum_{i=1}^{3} {\rm d} x^i {\rm d} x^i  ~. \label{charged}
\eeqn
where \cite{Hartnoll:2009sz}
\beqn
 \gamma(z) =  1 - \left( \frac{ z}{z_0}\right)^4 . \label{gammach}
\eeqn
We have introduced the tilde for the coordinate $\tilde t$ for later convenience. 
The length scale $L$ is the AdS radius and inversely related to the cosmological constant. 
The analog gravitational system will have to reproduce the metric (\ref{charged}) on a space-like
slice of dimension $3+1$ perpendicular to the horizon. Since the metric (\ref{charged}) is translationally
invariant into the directions parallel to the horizon, this just means that for the effective metric the sum in the last term runs
only over $1$ and $2$.

\section{Gauging the metric}

We now have to find a transformation that brings the metric (\ref{charged}) into the form (\ref{acmet}). As
noted in \cite{Hossenfelder:2014gwa}, this procedure generally isn't unique and one can spend a lot of time changing
coordinate systems in AdS space just to then realize that the resulting acoustic metric cannot be derived from any Lagrangian.
For this reason we will stay as close as possible to the transformation that was found to work previously and use the coordinate transformation $ \tilde t \to  t = \tilde t - f(z)$ also used earlier, but now
add the rescaling $z \to \tilde z = g^{-1}(z)$. This transformation does not change the $1/z^2$ prefactor of the
AdS metric, except that now $z$ is implicitly a function of $\tilde z$. We therefore can read off 
\beqn
\frac{\partial {\cal L}}{\partial \chi} = \frac{c L^2}{z^2}~\label{pref}
\eeqn
directly by comparing (\ref{acmet}) with (\ref{charged}).

Further comparing the components of the metric in the new coordinates with (\ref{acmet}) one obtains
\beqn
g_{tt} &:& - \gamma = -1 + (1-c^2) u_t u_t ~, \label{trafo1} \\
g_{\tilde z \tilde z} &:& \left(-\gamma (f')^2 + \frac{1}{\gamma}\right) (g')^2 = 1 + (1-c^2)u_{\tilde z} u_{\tilde z}~, \label{trafo2}\\
g_{t \tilde z} &:& - \gamma f' g'  = (1-c^2) u_t u_{\tilde z} ~, \label{trafo3}
\eeqn
where a dash denotes a derivative with respect to $z$. We have assumed here that like the metric of the planar black hole the condensed
matter system too is static and translationally invariant in the directions parallel to the horizon, so that the components of the velocity in the $x_i$-directions
all vanish. 

We now have three equations (\ref{trafo1},\ref{trafo2},\ref{trafo3}) for four unknown variables $f',g',c$ and one of the components of the four-velocity.
This does not look too promising but luckily with simple algebraic manipulations one can solve this system to
\beqn
c = g'~,~u_t^2 = \frac{z^4}{z_0^4(1- c^2)}~,~u_{\tilde z}^2 = \frac{c^2 - \gamma}{1-c^2}~, \label{rels}
\eeqn
and $f'$ can then be obtained by integration of
\beqn
f' =  \frac{\sqrt{(c^2-\gamma)(1-\gamma)}}{\gamma c}~.\label{f'}
\eeqn

To get the non-relativistic limit that reproduces the case discussed in \cite{Hossenfelder:2014gwa}, one needs $u_t^2 \to 1$, $u_{\tilde z}^2 \to z^4$ and
and $\tilde z \to z$. 
This tells us that the speed of sound must have an expansion of the form 
\beqn
c^2 = \gamma + a \left( \frac{z}{z_0} \right)^8 + {\cal O}(z^9)~, \label{cexp}
\eeqn 
where $a$ is a dimensionless
coefficient that we will determine in the next section.

\section{Checking that the equations of motion are fulfilled}

So, encouragingly we have seen that there is a straight-forward way to bring the metric of the planar black
hole into the right form for the acoustic metric. Next we will have to check whether a fluid with the degrees of freedom
identified in the previous section fulfills the equations of motion. As was noted in \cite{Hossenfelder:2014gwa}, the
Euler-equation doesn't actually give a constraint on the system, it instead determines the potential  necessary to 
obtain the right pressure gradient. The remaining consistency requirement is that the continuity equation be fulfilled,
which most conveniently can be written as the $t$-component of stress-energy conservation:
\beqn
\partial_\nu T^{\nu t} = 0~.
\eeqn

Since our system is time-independent and translationally invariant into the directions perpendicular to $z$ and the metric is
diagonal, this
simplifies to
\beqn
\partial_{z} \left( (\rho_0 + p_0) u^t u^{\tilde z} \right) = \partial_z \left( \frac{\partial {\cal L}}{\partial \chi} \chi u^t u^{\tilde z} \right) = 0~. \label{cont}
\eeqn
Because $\tilde z$ is just a rescaling $z$ and $c \partial_z = \partial_{\tilde z}$, it does not matter whether we take the derivative in this
equation with respect to $z$ or $\tilde z$.

The key to demonstrating that the continuity equation is fulfilled is to look at the equation of motion of the background field $\theta_0$. We will now, crucially, assume that the field $\theta_0$ describes the phase of some other field $\Phi_0 = \phi_0 \exp({\rm i} \theta_0)$. Then the kinetic term $\chi$ has only mass dimension two, and
there must be a prefactor of mass dimension two in the metric that comes from the amplitude of the field $\phi_0$. We will denote this prefactor with $m^2$ so that now
\beqn
\frac{\partial {\cal L}}{\partial \chi} = \frac{c L^2}{z^2} m^2~.\label{prefm}
\eeqn

The assumption that $\theta_0$ is a phase is sufficient to arrive at a metric that fulfils all requirements, but it might not be the only choice. It is chosen here because in the non-relativistic limit it reduces to the case earlier discussed in \cite{Hossenfelder:2014gwa}. However, it is quite possible that there are other ways to obtain the correct relativistic metric. The purpose here is merely to demonstrate that it is possible to obtain the relativistic metric, not to show that this procedure is the unique way to obtain it.

If $\theta_0$ is a phase, this also means that the potential $V$ will typically be a function of $\phi_0^2$ (and, possibly, the coordinates), but be independent of $\theta_0$. The equation of motion of
$\theta_0$ then reads
\beqn
\partial_\mu \sqrt{- \eta} \eta^{\mu \nu} \frac{\partial {\cal L}}{\partial \chi} \partial_\nu \theta_0 = 0~. \label{thetaeom}
\eeqn
It is then further natural to assume that the time-dependence of $\theta_0$ is just $mt$.
Using (\ref{pref}), we can then solve (\ref{thetaeom}) with
\beqn
\partial_t \theta_0 = m~,~\partial_{\tilde z} \theta_0 =  m  \frac{z^2}{L^2c}~.
\eeqn

With this we get from (\ref{rels})
\beqn
u_0^2 = \frac{m^2}{\chi} = \frac{z^4}{z_0^4(1- c^2)}~, u_{\tilde z}^2 = \frac{m^2 z^4}{\chi c^2 L^4} = \frac{c^2 -\gamma}{1-c^2}~, \label{chi2}
\eeqn
and so
\beqn
\chi = m^2 \frac{z_0^4}{z^4} (1-c^2)~, \label{chi}
\eeqn
and
\beqn
c^2 (c^2 - \gamma) = \frac{z^8}{L^4 z_0^4} ~\Rightarrow~ c^2 = \frac{\gamma}{2} + \sqrt{ \frac{\gamma^2}{4} +  \left( \frac{z^2}{z_0 L}\right)^4}~,\label{c2}
\eeqn
where we have picked a sign so that $c^2 \to 1$ for $z \to 0$. With this solution for $c$ we could now integrate $f$ and $g$ via Eqs (\ref{rels}) and (\ref{f'}) but these
functions are not particularly illuminating. More interesting is that the speed of sound can be expanded as
\beqn
c^2 = 1 - \left( \frac{z}{z_0}\right)^4 + \left( \frac{z}{z_0}\right)^4 \bigg( \frac{z}{L}\bigg)^4 + {\cal O}(z^{12})~,
\eeqn
and thus has indeed the form anticipated in (\ref{cexp}).

We can then finally insert all these expressions into the continuity equation (factors appear in the same order as in (\ref{cont})):
\beqn
\partial_z \left( \left( m^2 \frac{c L^2}{z^2} \right) \left( m^2 \frac{z_0^4}{z^4} (1-c^2) \right) \sqrt{\frac{z^4 (c^2 - \gamma)}{z_0^4 (1-c^2)^2}} \right) &=& 0~.
\eeqn
and see that with (\ref{c2}) this equation is fulfilled. 

One convinces oneself easily that the reason
this equation is fulfilled has nothing to do with the particular form of the four-velocity or the prefactor of the metric. 
It is not the scaling of the 
components that is relevant here, but that
the $z$-coordinate is
chosen as the coordinate system in the laboratory in which the metric is of the Minkowski-form. If $z$
was a radial coordinate for example then this equation would not be fulfilled.

Now that we have a relativistic analog metric, we have to note however that
it does not have a non-relativistic limit! 
The reason is that the limit where $u_z \ll 1$ is the same limit in
which $c \to 1$. In the limit $c \to 1$ the acoustic metric (\ref{acmet}) seems to become ill-defined. However, one
can absorb the factors $(1-c^2)$ from the metric (\ref{acmet}) into
the four-velocity and then take the limit $c\to 1$. The metric then remains well-defined but the 
four-velocity becomes a null-vector, rendering the non-relativistic limit meaningless.

To avoid this, one 
can rescale the
time coordinate $t \to t \kappa$ with an arbitrary constant $\kappa$, which has the effect of
changing the limit of $c$ from $1$ to $\kappa$. The equations (\ref{trafo1},\ref{trafo2},\ref{trafo3}) then
become
\beqn
g_{tt} &:& - \gamma \kappa^2 = -1 + (1-c^2) u_t u_t ~, \label{trafo1a} \\
g_{\tilde z \tilde z} &:& \left(-\gamma (f')^2 + \frac{1}{\gamma}\right) (g')^2 = 1 + (1-c^2)u_{\tilde z} u_{\tilde z}~, \label{trafo2a}\\
g_{t \tilde z} &:& - \gamma f' g' \kappa  = (1-c^2) u_t u_{\tilde z} ~, \label{trafo3a}
\eeqn
and one finds similarly as before (compare to \ref{rels})
\beqn
c = \kappa g'~,~u_t^2 = \frac{1-\kappa^2 \gamma}{1- c^2}~,~u_{\tilde z}^2 =  \frac{c^2 - \kappa^2 \gamma}{1-c^2}~. \label{rels2}
\eeqn
The continuity equation is still fulfilled in this case, and the speed of sound now is approximately
\beqn
c^2 = \kappa^2 \gamma(z) + \left( \frac{1-\kappa^2}{\kappa^2}\right) \frac{z^4}{L^4} + {\cal O} (z^8)~.
\eeqn

The above calculation reveals that this finding that the continuity equation is fulfilled is independent of the number of
dimensions of the AdS-space of which our analog systems describes a spatial slice. This might seem surprising because 
it was found in \cite{Hossenfelder:2014gwa} that the non-relativistic limit
only works if the AdS-space is $4+1$ dimensional. How can this be? The reason can be found in equation
(\ref{trafo2}) or (\ref{trafo2a}) respectively. Once one has identified the
four-velocity it turns out the square of the
velocity $u_{\tilde z}$ is comparable to terms that appear also in the metric coefficients. This means that if one
starts with the non-relativistic metric, in which the term $u_{\tilde z}^2$ in Eq (\ref{trafo2}) or (\ref{trafo2a}) is just absent, one neglects
a relevant term and obtains a different $f'$ that leads to another coordinate system. If one identifies the degrees
of freedom in this different coordinate system, the
continuity equation is fulfilled only in $4+1$ dimensions. So the discrepancy comes about because
two different coordinate systems were used.

\section{Working towards a Lagrangian}

We have seen that the degrees of freedom of a fluid which can be extracted from the metric of
the planar black hole do fulfill the equations of motion. This leads to the question whether we
can find a Lagrangian for this system. From Eqs (\ref{sos}), (\ref{prefm}) and (\ref{chi2}) we can infer that this Lagrangian
must have the property that
\beqn
 \frac{\partial {\cal L}}{\partial \chi} = m^2 \frac{c L^2}{z^2}~,~\frac{\partial^2 {\cal L}}{\partial \chi^2} = \frac{z^2 L^2}{2cz_0^4}~. \label{req}
\eeqn

We can obtain a Lagrangian for example by making the ansatz
\beqn
{\cal L} = \frac{2}{3} m \left( \chi - V/m^2 \right)^{3/2}~, \label{lans}
\eeqn
where the constant $m$ is introduced for dimensional reasons.
Lagrangians with fractional powers seem strange from the perspective of particle physics, but they are a common
occurrence in effective theories for condensed matter systems \cite{Son:2002zn}. The reason is, roughly speaking, that to obtain the
effective Lagrangian one uses a limit in which part of the equations of motion can be solved and then substitutes this
solution back into the Lagrangian. Since this partial solution itself depends generically on the kinetic 
term $\chi$ this can lead to
the odd-looking fractional powers. Interestingly enough, 
Lagrangians with the particular power $3/2$ of the
kinetic term have recently been proved useful in the cosmological context, see for example \cite{Berezhiani:2015bqa} and references
therein.  

The ansatz (\ref{lans}), when compared with (\ref{req}) leads to
\beqn
{\cal L} = \frac{2}{3}m \left( \chi - m^2 z_0^4/z^4 \right)^{3/2}~. \label{lagex}
\eeqn
Again the potential term might appear somewhat unmotivated, but keep in mind that this is
an effective theory. It might come about for example from a Lagrangian that originally was of the form
\beqn
{\cal L} = \frac{2}{3} \left[ \left( \eta^{\mu \nu } (\partial_\nu \Phi_0) (\partial_\mu \Phi_0^*)/(\Phi \Phi^*) -  \tilde{V}  z_0^4/z^4  \right) \right]^{3/2}~,~ 
\tilde{V} = 2 \Phi \Phi^* - \frac{(\Phi \Phi^*)^2}{m^2} \label{lagphi}
\eeqn
where $\Phi_0 = \phi_0 \exp {\rm i} \theta_0$. One can then decompose the kinetic term of $\Phi_0$ as
\beqn
\eta^{\mu \nu } (\partial_\nu \Phi_0) (\partial_\mu \Phi_0^*)  =  \eta^{\mu \nu } (\partial_\nu \phi_0) (\partial_\mu \phi_0) + \phi^2 \chi~,
\eeqn
and derive the equation of motion for $\phi_0$
\beqn
\partial_{\tilde z} \frac{1}{\phi_0^2} \frac{\partial {\cal L}}{\partial \chi} \partial_{\tilde z} \phi_0 -  \frac{\partial {\cal L}}{\partial \chi} \frac{1}{2} \frac{(\partial_{\tilde z} \phi_0)^2}{ \phi^3_0} + \frac{\partial {\cal L}}{\partial \chi} \frac{{\rm d} \tilde V}{{\rm d} \phi_0} =0~.
\eeqn

This equation of motion has a simple solution in which the field $\phi_0$ is just constant and sits in the potential minimum where
\beqn
\phi_0^2 = \tilde V = m^2~.
\eeqn
If one inserts this solution back into the Lagrangian (\ref{lagphi}) one obtains (\ref{lagex}). This is just to demonstrate
how this type of potential can arise. It would be nicer of course
if one could find a background that creates the $z$-dependence rather than putting this into the potential to begin with.
And besides this, while the Lagrangian (\ref{lagphi}) looks more plausible than (\ref{lagex}) it would be interesting to see whether the Lagrangian (\ref{lagphi}) itself can be obtained as the low-energy
limit from some other theory, but we will leave this to a future work. For now it is good to have an effective Lagrangian 
at all, because this allows one to look for other solutions to the field equations.

\section{Summary}

We have shown here that a $3+1$ dimensional slice of the metric of the planar black hole in AdS-space of arbitrary dimension can
be obtained as an effective acoustic metric, and that a simple effective Lagrangian exists for this system which has a fractional
power of $3/2$ in the kinetic term.



\end{document}